\documentclass[conference]{IEEEtran}
\IEEEoverridecommandlockouts

\usepackage{algorithm}
\usepackage{algorithmic}
\usepackage[geometry]{ifsym}
\usepackage{ifsym}
\usepackage{amssymb}
\usepackage{yhmath}
\usepackage{amscd}
\usepackage{dsfont}
\usepackage{amsmath}
\usepackage{epsfig}
\usepackage{graphics}
\usepackage{psfrag}
\usepackage{rotating}
\usepackage{amsmath}
\usepackage{amsfonts}
\usepackage{url}
\usepackage{color}
\usepackage{epstopdf}
\usepackage[printonlyused]{acronym}
\newtheorem{lemma}{Lemma}
\newtheorem{theorem}{Theorem}
\newtheorem{definition}[theorem]{Definition}
\newtheorem{proposition}{Proposition}
\usepackage{mdwlist}%tight lists
\usepackage[draft,footnote,nomargin]{fixme}
\definecolor{orange}{rgb}{1,0.5,0}

\newcommand{\sref}[1]{Sec.~\ref{#1}}

\begin{document}

%\title{Energy Efficient Power Allocation with Non-Cooperative Mobile Users}
%\title{Energy-Efficient Power Allocation with Non-Cooperative Channel State Feedback in Cellular Networks}
\title{An Energy-Efficient Power Allocation Game with Selfish Channel State Reporting in Cellular Networks}

%\author{Fran\c cois M\'eriaux$^{+\ddagger}$\thanks{Fran\c cois M\'eriaux has been supported by the \emph{PhD@Bell Labs} internship program.}, Stefan Valentin$^\ddagger$, Samson Lasaulce$^+$, and Michel Kieffer$^+$\thanks{Michel Kieffer has been partly supported by the Institut Universitaire de France.}}

%\medskip\\
% 	\email{meriaux@lss.supelec.fr, stefan.valentin@alcatel-lucent.com,}\\
% 	\email{lasaulce@lss.supelec.fr, kieffer@lss.supelec.fr}\medskip\\
% 	\affaddr{$^+$Laboratoire des Signaux et Syst\`emes -- LSS (CNRS-SUPELEC-Paris Sud), Gif-sur-Yvette, France}\\
% 	\affaddr{$^\ddagger$Bell Labs, Alcatel-Lucent Deutschland AG, Stuttgart, Germany}
% }

\author{\IEEEauthorblockN{Fran\c cois M\'eriaux\IEEEauthorrefmark{1}\IEEEauthorrefmark{2}\thanks{Fran\c cois M\'eriaux has been supported by the \emph{PhD@Bell Labs} internship program.},
Stefan Valentin\IEEEauthorrefmark{2},
Samson Lasaulce\IEEEauthorrefmark{1} and
Michel Kieffer\IEEEauthorrefmark{1}\thanks{Michel Kieffer has been partly supported by the Institut Universitaire de France.}}
\IEEEauthorblockA{\IEEEauthorrefmark{1}Laboratoire des Signaux et Syst\`emes -- LSS (CNRS-SUPELEC-Paris Sud), Gif-sur-Yvette, France \\
Email: \{meriaux,lasaulce,kieffer\}@lss.supelec.fr}
\IEEEauthorblockA{\IEEEauthorrefmark{2}Bell Labs, Alcatel-Lucent Deutschland AG, Stuttgart, Germany\\
Email: stefan.valentin@alcatel-lucent.com}}

% \author{\IEEEauthorblockN{Michael Shell}
% \IEEEauthorblockA{School of Electrical and\\Computer Engineering\\
% Georgia Institute of Technology\\
% Atlanta, Georgia 30332--0250\\
% Email: http://www.michaelshell.org/contact.html}
% \and
% \IEEEauthorblockN{Homer Simpson}
% \IEEEauthorblockA{Twentieth Century Fox\\
% Email: homer@thesimpsons.com}
% \and
% \IEEEauthorblockN{James Kirk\\ and Montgomery Scott}
% \IEEEauthorblockA{Starfleet Academy\\
% San Francisco, California 96678-2391\\
% Telephone: (800) 555--1212\\

\maketitle
\begin{abstract}
%Energy-efficient ressource allocation is a powerful approach to reduce the operation costs and environmental footprint of cellular networks. 
With energy-efficient resource allocation, mobile users and base station have different objectives. While the base station strives for an energy-efficient operation of the complete cell, each user aims to maximize its own data rate. To obtain this individual benefit, users may selfishly adjust their \ac{CSI} reports, reducing the cell's energy efficiency. To analyze this conflict of interest, we formalize energy-efficient power allocation as a utility maximization problem and present a simple algorithm that performs close to the optimum. By formulating selfish CSI reporting as a game, we prove the existence of an unique equilibrium and characterize energy efficiency with true and selfish CSI in closed form. Our numerical results show that, surprisingly, energy-efficient power allocation in small cells is more robust against selfish CSI than cells with large transmit powers. This and further design rules show that our paper provides valuable theoretical insight to energy-efficient networks when CSI reports cannot be trusted. 

%What happens with the overall energy-efficiency of power allocation when users are allowed to alter their reported CSI values?
\end{abstract}

%A category including the fourth, optional field follows...
%\category{D.2.8}{Software Engineering}{Metrics}[complexity measures, performance measures]


\begin{keywords}Energy Efficiency, Power Allocation, Feedback, Game Theory \end{keywords}

\section{Introduction}
Reducing the energy consumption of cellular networks is a challenging task for network operators and telecommunication equipment vendors. One relevant approach is the energy-efficient allocation of wireless resources such as bandwidth, time and transmit power. By carefully allocating these resources to the mobile users, a base station can reduce the energy consumption of the complete cell \cite{tan09:energy_robustness,KSF10trcf}.

However, such centralized form of energy-efficient resource allocation raises two problems. The first problem results from the time-variant nature of the wireless fading channel. To adapting to the users' varying channels, the base station has to update the resource allocation frequently, e.g., once per millisecond in most LTE systems \cite[Fig. 5.1-1]{3gpp10:sys_descr}. This renders computational complex approaches infeasible. The second problem results from fading and interference, causing most wireless channels to be non-reciprocal. Consequently, the channel state is only known at the mobile user and needs to be transferred to the base station. The accuracy of this \ac{CSI} feedback is crucial for the quality of the resource allocation decision. If users report inaccurate \ac{CSI}, the resource allocation does not reflect the actual channel states and the performance of the overall network may degrade.

In this paper, we analyze energy-efficient resource allocation when users and base station have different interests. While the base station strives for energy efficiency of the whole cell, each users aims to maximize its individual \ac{SNR}. Note that in a perfect cellular network the base station would simply overrule the users' interest. In practice, however, the base station controls the CSI format \cite[Sec. 7.2]{hsdpa09:phy_spec} but not how mobile users generate the reported CSI values. Thus, mobile users can selfishly report CSI that provides them an individual benefit while the performance of the overall network suffers. 

This makes it necessary to analyze the effect of selfishly chosen CSI on the performance of the wireless cell. To do so, we apply a game theoretical approach. Here, the mobile users are reflected as players who selfishly choose their CSI to maximize their individual performance. For all users, the base station computes the resource allocation by maximizing a utility function. This function expresses the energy efficiency of the complete cell. Limiting our analysis to power allocation allows us to formally compare centralized power allocation with true CSI to the allocation with selfishly chosen CSI. Consequently, our results provide significant insight to energy-efficient cellular networks in which the CSI accuracy cannot be trusted. 

%Im 3GGP Standard 3GPP TS 36.213, ist im Kapitel “7.2.3 Channel quality indicator (CQI) definition” eine Tabelle (Table 7.2.3-1: 4-bit CQI Table) mit den 16 CQI Werten gegeben, welche von Terminal zur Basisstation gesendet werden können.
%
%Aperiodic CQI/PMI/RI reporting on PUSCH (36.213, 7.2.1),
%
%Für das „Periodic CQI/PMI/RI reporting on PUCCH [36.213, 7.2.2]” sind folgende Reporting-Perioden definiert:  NP = {2, 5, 10, 20, 32, 40, 64, 80, 128, 160}  subframes.
%
%3GPP TS 36.300 in Kapitel 5:
%
%Physical Layer for E-UTRA
%
%Frame structure Type 1 is illustrated in Figure 5.1-1. Each 10 ms radio frame is divided into ten equally sized sub-frames.

\subsection{Contributions}
In particular, we make the following contributions:
\begin{enumerate*}
\item We analyze the optimization problem, where the base station maximizes the cell's energy efficiency under power constraints. We obtain an optimality condition and propose a simple power allocation algorithm whose energy efficiency is close to the optimum in general.
\item We prove that there exists a unique equilibrium for the users' selfish choice of the CSI reports. This allows studying the network in a stable state, where no user has an unilateral interest to report a different CSI.
\item We compare the performance with true and selfish CSI reporting. For both types of reports, we provide a closed-form result for the cell's energy efficiency. Numerical results surprisingly show that small cells are more robust to selfish CSI than cells with large transmit power constraints.
\end{enumerate*}
All in all, our paper provides the theoretical insight to cope with selfish CSI reports in cellular networks with energy-efficient power allocation.

\subsection{Paper Structure}
Our paper is structured as follows. After discussing the novelty of our study with respect to related work in \sref{sec:relwork}, we describe the studied cellular system in \sref{sec:sysmod_scen}. Energy efficiency is formalized in terms of a constrained utility maximization problem in \sref{sec:sysmod_util}. Using this function, we formulate the non-convex optimization problem in \sref{sec: EEschedu}. Having discussed further properties of this problem, we provide an optimality condition and derive a power allocation algorithm. \sref{sec:game_study} is devoted to the game theoretical study of selfish CSI reports. Therein, we prove the existence of an unique equilibrium for the proposed energy-efficient power allocation. Finally, the numerical results in \sref{sec: Numeric} point out by how much a selfish choice of CSI reduces the energy efficiency of the system and by how much it increases the users' individual \acp{SNR}.

\section{Related Work and Novelty}\label{sec:relwork}
Our paper joins two fields of research. The first field is energy-efficient power control, where often transmit power is minimized under \ac{QoS} constraints \cite{Foschini-vt-1993,Yates-jsac-1995}. We do not follow this common approach in our work. Instead, we maximize of the ratio between throughput and transmit power. This utility function was introduced in \cite{goodman-pc-2000} and was used in \cite{meshkati-jsac-2006,buzzi-jstp-2011} for \ac{CDMA} systems. With CDMA, employed such utility function for power allocation results in a game where the players interact via the interference term. Our work differs, as the players interact via the power allocation algorithm of the base station. Such centralized resource allocation is common with cellular systems such as LTE and has not been studied with the utility functions from \cite{goodman-pc-2000} so far.

The second related field is scheduling with non-cooperative mobile users. Typically, a centralized scheduler at the base station allocates wireless channel resources to maximize the instantaneous sum throughput of its cell. As such allocation is based on CSI reports, some users may not report their actual channel gains to obtain more resource from the base station. In \cite{Kavitha2012}, Kavitha et al. model this selfish choice of CSI as a signaling game \cite{Sobel2009}. Here, even the base station is considered as a player that knows which mobile users cooperate and which not. The same authors take a similar approach in \cite{Kavitha2010} for $\alpha$-fair scheduling. Here, a signaling game cannot be used since each resource allocation is affected by the scheduling history. 

Although we focus on selfish CSI reporting, our work differs significantly. Unlike the above papers, we do not include the base station in the set of players. Instead, the base station performs centralized power allocation under its own general objective -- energy efficiency per cell. This objective differs from the users' aim, which strive for maximizing their individual SNR. This conflict of interest between the user's individual objective and the base station's objective per cell has not been studied so far.

\section{System Model}\label{sec:sysmod}
\subsection{Wireless Scenario}\label{sec:sysmod_scen}
We study a single wireless cell where one base station allocates transmit power to $K$ mobile users during the downlink. An arbitrary user is denoted by $k \in \mathcal{K}= \{1,\ldots,K\}$ and the downlink transmit power is $p_k \in [0, P]$ Watts for each user. Time is divided into slots and power allocation is done once per slot. To focus on the effect of power allocation and to provide tractable results we ignore subband and time slot allocation. Consequently, each user has its own, fixed subband of bandwidth $W$ and inter-cell interference is ignored. 

For each subband, the wireless channel from the base station to the user is assumed to experience quasi-static, time-selective fading with channel coefficient $h_k$. The fading process is assumed to be i.i.d.\ Rayleigh, which leads to exponentially distributed channel gains $|h_k|^2$. For each mobile user $k\in \mathcal{K}$, we can write the instantaneous SNR as
\begin{equation}
 \gamma\bigl(p_k,|h_k|^2\bigr) = \frac{p_k |h_k|^2}{\sigma^2},
\end{equation}
with $\sigma^2$ the variance of the noise for user $k$. To perform power allocation, the base station requires a CSI report from every mobile user per time slot. In this work, we require that this feedback is equivalent to the channel gain $|h_k|^2$, $\forall k \in \mathcal{K}$.

%In this work, we consider that this feedback is $\mathbb{E}[|h_k|^2]$, $\forall k \in \mathcal{K}$, i.e., the expectation of the channel gain for the current time slot. Considering that both mobiles users and the base station have knowledge of the expected channel gains only, the averaged SNR is given by
%\begin{equation}
% \gamma\bigl(p_k,\mathbb{E}[|h_k|^2]\bigr) = \frac{p_k \mathbb{E}[|h_k|^2]}{\sigma^2}.
%\end{equation}
%From a practical viewpoint, the expectation of the channel gain can be estimated by a moving average over the $\tau$ previous channel gains realizations, with $\tau \in \mathbb{N}$. For each mobile transmitter, the expected value may be different from a time slot to another. 

\subsection{Energy-Efficient Utility of the Cell}\label{sec:sysmod_util}
The base station performs power allocation, taking into account the energy efficiency associated with each of the mobile users. There are many common uses of the term energy efficiency. Here, we precisely refer to the notion introduced by Goodman and Mandayam in \cite{goodman-pc-2000}, i.e., the energy efficiency is the ratio between the effective throughput of the mobile user and the transmit power spent to attain this throughput. Contrary to \cite{goodman-pc-2000} in which power control is considered for the uplink, we perform power allocation in the downlink. Hence, the transmit power considered in the energy efficiency is not the power from the mobile user but the power allocated to the mobile user by the base station. We denote $u_{BS}^{(k)}$ the energy-efficient utility associated with the $k^{\text{th}}$ mobile user. It writes
\begin{equation}
 u_{BS}^{(k)}(p_k) = \frac{R f\bigl(\gamma_k\bigl(p_k,|h_k|^2\bigr)\bigr)}{p_k},
\label{eq:ee_per_usr}
\end{equation}
with $R$ being a fixed transmission rate of the base station in bit/s, and $f$ is an S-shaped function taking its values in $[0,1]$ which represents the packet success rate during transmission. This function depends on the expected SNR of each mobile user $k \in \mathcal{K}$. The power $p_k$ is expressed in Watt, the unit of the utility is hence bit/Joule. Similarly to Belmega and Lasaulce work in \cite{belmega-tsp-2011}, for the particular case of SISO channel, the efficiency function $f$ is defined as
\begin{equation}
\begin{aligned}
  f\bigl(\gamma_k\bigl(p_k,|h_k|^2\bigr)\bigr) &= 1 - P_{\text{out}}\bigl(\gamma_k\bigl(p_k,|h_k|^2\bigr),a\bigr), \\
&= \exp\biggl(-\frac{a}{\gamma_k(p_k,|h_k|^2)}\biggr),
\end{aligned}
\end{equation}
with $a=2^{R/W}-1$ as the threshold under which the SNR causes an outage. It is important to notice that there is an approximation here, as $P_{\text{out}}$ should depend on the expectation of the channel gain, and not on the instantaneous value of the channel gain. The justification of this approximation is that power allocation is performed every time slot. Consequently, instantaneous channel gains are required to take into account the variations of the characteristics of the channels from one time slot to another. 

\begin{figure}
\centering
\includegraphics[width=\linewidth]{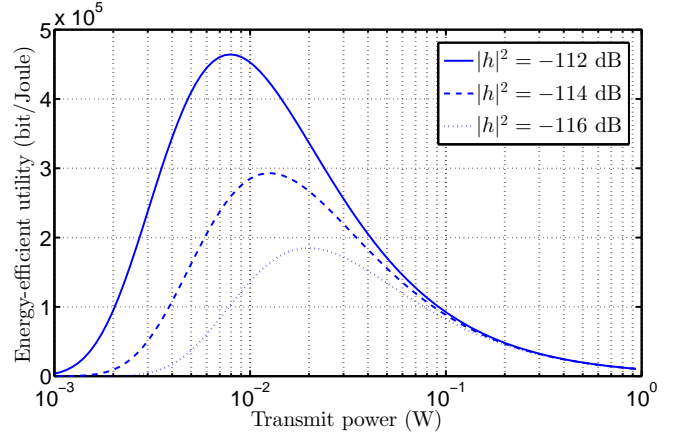}
\caption{Energy-efficient utility per user with respect to the allocated power for three different channel gains. Parameters are $a=1$, and $\sigma^2 = 5\times 10^{-14}$ W.}
\label{fig: EEuti}
\end{figure}

Fig.~\ref{fig: EEuti} illustrates the typical shapes of this energy-efficient utility per user, for three different values of the channel gain. It represents the energy-efficient utility associated with one mobile user (in bit/Joule) with respect to the power allocated to that user. Energy-efficient utility maximization does not necessarily correspond to rate maximization. Indeed, contrary to a rate maximization, it occurs that it is not always optimal to use all the power available to maximize energy efficiency. It is also interesting to note that for the same transmit power, the energy-efficient utility per user increases with the channel gain. Regarding individual optimal power, we can check that the energy-efficient utility per user can be maximized with a lower power when the channel gain is higher. For a cellular energy-efficient power allocation perspective, it means that it is more interesting to allocate power to mobile users with a good channel gain, as these mobile users require less power and offer a better energy-efficient utility.

The energy-efficient utility per cell is the sum of all the energy-efficient utilities per user, i.e.,
\begin{equation}\label{eq: uti_bs}
 u_{BS}(\mathbf{p}) = \sum_{k \in \mathcal{K}} u_{BS}^{(k)}(p_k),
\end{equation}
where the vector $\mathbf{p}$ contains all power values allocated by the base station. 

%This expression is preferred to the ratio between the sum of all the effective throughputs and the sum of the powers for optimal power allocation considerations. Indeed, such a utility would lead to allocate no power to some mobile users even without total power constraint whereas with (\ref{eq: uti_bs}), the optimal power allocation consists in allocating power to all the mobile users (without power constraint). Hence, the considered utility is fairer for mobile users.

Applying these utility functions for optimal power allocation requires to understand their properties. For each user, energy efficiency is expressed by (\ref{eq:ee_per_usr}). This function $u_{BS}^{(k)}(p_k)$ is continuous with respect to $p_k$. Consequently, the sum (\ref{eq: uti_bs}) is continuous with respect to $\mathbf{p}$. Nonetheless, such similarity cannot be found for another property of the functions: although each of the individual functions $u_{BS}^{(k)}(p_k)$ is quasiconcave, their sum is neither concave nor quasiconcave. 
%This property of (\ref{eq: uti_bs}) is illustrated in Fig.~\ref{fig: 2players}.

% However, $|h_k|^2$ being an exponentionnaly distributed random variable~:
% \begin{equation}
% \begin{aligned}
%  P_{\text{out}}(\gamma_k(p_k,\mathbb{E}[|h_k|^2]),a) &= Pr\biggl[SNR(p_k,|h_k|^2) < a\biggr], \\
% &= Pr\biggl[|h_k|^2 < \frac{\sigma^2 a}{p_k}\biggr], \\
% &= \int_0^{\frac{\sigma^2 a}{p_k}} \frac{1}{\mathbb{E}[|h_k|^2]} \exp\biggl(-\frac{x}{\mathbb{E}[|h_k|^2]}\biggr) \text{d}x, \\
% &= 1 - \exp\biggl(-\frac{\sigma^2 a}{p_k \mathbb{E}[|h_k|^2]}\biggr), \\
% &= 1 - \exp\biggl(-\frac{a}{\gamma_k\bigl(p_k,\mathbb{E}[|h_k|^2]\bigr)}\biggr).
% \end{aligned}
% \end{equation}
% Then 
% \begin{equation}
%  f\bigl(\gamma_k\bigl(p_k,\mathbb{E}[|h_k|^2]\bigr)\bigr) = \exp\biggl(-\frac{a}{\gamma_k(p_k,\mathbb{E}[|h_k|^2])}\biggr).
% \end{equation}

\section{Energy-Efficient Power Allocation}\label{sec: EEschedu}
\subsection{Optimization Problem}
The objective of the base station is to allocate at most $P$ Watts among the different mobile users in order to maximize the energy-efficient utility per cell (\ref{eq: uti_bs}). The optimization problem is
\begin{equation}\label{eq: opti_pb}
 \begin{aligned}
&\underset{\mathbf{p} \in \mathbb{R}^K}{\text{maximize}} & & \sum_{k \in \mathcal{K}} u_{BS}^{(k)}(p_k), \\
&\text{subject to } & & \sum_{k \in \mathcal{K}} p_k \leq P, \\
& & & p_k \geq 0,\; \forall k \in \mathcal{K}.
 \end{aligned}
\end{equation} 
The problem is solved by choosing the optimization variables $p_1,\ldots,p_k$ such that the $K+1$ constraints hold. These constraints ensure that the optimization set is compact convex. The energy-efficient utility per cell $u_{BS}$ being continuous with respect to $\mathbf{p}$, we know that $u_{BS}$ has at least one maximum in the set delimited by the power constraints. Hence, there exists at least one optimal solution to this problem. But due to the utility $u_{BS}$ being not quasiconcave, this optimization problem is hard to solve. 

\subsection{Optimality Conditions}\label{sec: slopes_prop}
Optimality conditions can be obtained by the study of the Lagrangian associated with (\ref{eq: opti_pb}). It writes 
\begin{equation}
\mathcal{L}(\mathbf{p},\lambda,\mathbf{\mu}) = \sum_{k \in \mathcal{K}} u_{BS}^{(k)}(p_k) - \lambda \biggl(\sum_{k \in \mathcal{K}} p_k - P\biggr) + \sum_{k \in \mathcal{K}} \mu_k p_k.
\end{equation}
The optimality conditions write $\forall k \in \mathcal{K}$
\begin{equation}
\frac{ \text{d} u_{BS}^{(k)}(\bar{p}_k)}{\text{d} p_k} = \lambda - \mu_k,
\end{equation}
with $\mathbf{\bar{p}}$ an optimal power allocation.
%Hence, we can derive necessary conditions for an optimal energy-efficient power allocation.
These optimality conditions lead to a partition of the set $\mathcal{K}$ in $\mathcal{K}'$ and $\mathcal{K}''$. The set $\mathcal{K}'$ is the subset of users for which power is allocated, leading to $\mu_k =0$ and $\frac{ \text{d} u_{BS}^{(k)}(p_k)}{\text{d} p_k} = \lambda$. The set $\mathcal{K}''$ is the subset of users with no power allocated ($\bar{p}_k=0$), for which $\mu_k=\lambda$.
% \begin{proposition}\label{prop: slopes_prop}
% Let $\mathbf{\bar p}$ be an optimal energy-efficient power allocation. There exists a subset $\mathcal{K}' \subseteq \mathcal{K}$ such that $\forall i, j \in \mathcal{K}'$
% \begin{equation}\label{eq:slopes_prop}
% \frac{\text{d} u_{BS}^{(i)}(\bar{p}_i)}{\text{d} p_i} = \frac{\text{d} u_{BS}^{(j)}(\bar{p}_j)}{\text{d} p_j},
% \end{equation}
% and $\forall k \in \mathcal{K}^{''} =  \mathcal{K}\setminus \mathcal{K}'$
% \begin{equation}
% \bar{p}_k = 0.
% \end{equation}
% \end{proposition}
In other words, in an optimal energy-efficient power allocation, there is a slope equality condition for a subset of the users, and the remaining users are given no power.

%If the optimal $p_k$ is different from $0$, then $\mu_k = 0$, and we have a slope equality. If the optimal $p_k$ is equal to zero, necessarily $\mu_k = \lambda$.

\subsection{Algorithm Design Principle}\label{sec: idea}
Here we do not provide an optimal power allocation scheme to solve (\ref{eq: opti_pb}). Instead, we propose a simple suboptimal algorithm with a performance that is very close to the optimum. 
%illustrated in Fig. \red{[Figure to be added]}. 
If the sum power constraint is not saturated, $\lambda = 0$, $\mathcal{K}'  = \mathcal{K}$. Then, the proposed algorithm even provides the optimal allocation per cell. 

The idea behind the proposed algorithm is that, without the sum power constraint, (\ref{eq: opti_pb}) can be divided into $K$ simple quasiconcave optimization problems. For each of these $K$ problems, the individual optimal power is known. As given in \cite{belmega-tsp-2011}, $\forall k \in \mathcal{K}$, the power $p_k^*(|h_k|^2)$ that maximizes the energy-efficient utility of mobile user $k$ is
\begin{equation}\label{eq: indi_EE_pow}
\begin{aligned}
 p_k^*(|h_k|^2) &= \arg \max_{p_k} \frac{f\bigl(\gamma_k\bigl(p_k,|h_k|^2\bigr)\bigr)}{p_k}, \\
&= \min\biggl\{\frac{\sigma^2 a}{|h_k|^2},P\biggr\}.
\end{aligned}
\end{equation}
This power is called individual optimal power. If the sum of all these individual optimal powers is less than the sum power constraint $P$, expression (\ref{eq: indi_EE_pow}) provides the solution of the optimization problem . If the sum exceeds $P$, the base station cannot allocate the individual optimal power $p_k^*$ for each mobile user. Then it has to choose which mobile users to serve and which users to exclude in order to maximize the energy-efficient utility per cell while satisfying the sum power constraint. How to make this choice is justified by Lemma \ref{lemma: 1}.

\begin{lemma}\label{lemma: 1}
$\forall i,j \in \mathcal{K},\; i \neq j,\; |h_i|^2 \geq |h_j|^2$ is equivalent to
\begin{equation}
p_i^*(|h_i|^2) \leq p_j^*(|h_j|^2),
\end{equation}
and
\begin{equation}
u_{BS}^{(i)}\bigl(p_i^*(|h_i|^2),|h_i|^2\bigr) \geq u_{BS}^{(j)}\bigl(p_j^*(|h_j|^2),|h_j|^2\bigr) .
\end{equation}
\end{lemma}
This means that a high individual  optimal power offers a poor outcome in terms of energy-efficient utility, whereas a low individual  optimal power offers a good outcome. Hence, from the base station perspective, it is more interesting to allocate the power budget for the users with the lower individual optimal power values. Note that contrary to water-filling, the individual optimal power associated with a mobile user with a good channel gain is less than the individual optimal power associated with a mobile user with a low channel gain. This is due to the fact that energy efficiency is maximized instead of sum capacity.

\subsection{Algorithm}
Based on the observations of Sec.~\ref{sec: idea}, we propose the following algorithm to allocate power in the cell, assuming that the coefficients provided by (\ref{eq: indi_EE_pow}) $\bigl(p_1^*(|h_1|^2),\ldots,p_K^*(|h_K|^2)\bigr)$ are in increasing order (to simplify the notation, we only write $(p_1^*,\ldots,p_K^*)$ in what follows). The power allocated to mobile user $k$ by the base station with that particular algorithm is denoted by $\tilde{p}_k$.
%\begin{algorithm}
%\caption{Process $\tilde{p}$}
%\begin{algorithmic}
%\REQUIRE $\bigl(p_1^*(\mathbb{E}[|h_1|^2]),\ldots,p_K^*(\mathbb{E}[|h_K|^2])\bigr)$, $P$.
%\STATE $\tilde{\underline{p}} = \underline{p}^*$
%\IF{$\sum_{k \in \mathcal{K}} \tilde{p}_k > P$} 
%\STATE $k' = K$
%\WHILE{$\sum_{k=1}^{K} \tilde{p}_k > P$}
%\STATE $\tilde{p}_{k'} \gets 0$
%\STATE $k' \gets k'-1$
%\ENDWHILE
%\STATE $\tilde{p}_{k'+1} = P-\sum_{k=1}^{k'} \tilde{p}_k$
%\ENDIF
%\RETURN $\tilde{p}$
%\end{algorithmic}
%\end{algorithm}

\begin{algorithm}
\begin{algorithmic}[1]\label{algo: EE_scheduling}
\caption{}
\REQUIRE $(p_1^*,\ldots,p_K^*)$, $P$.
\STATE $(\tilde{p}_1,\dots,\tilde{p}_K) = (0,\ldots,0)$
\STATE $i \gets 1$
\WHILE{$\sum_{k=1}^{K} \tilde{p}_k < P$}
\IF{$p_i^* < p_{i+1}^*$}
\IF{$\sum_{k=1}^{i} p^*_k < P$}
\STATE $\tilde{p}_i \gets p_i^*$
\ELSE 
\STATE $\tilde{p}_i \gets P - \sum_{k=1}^{i-1} \tilde{p}_k$
\ENDIF
\STATE $i \gets i+1$
\ELSE 
\STATE $j \gets i$
\WHILE{$p_i^* == p_{i+1}^*$}
\STATE $i \gets i+1$
\ENDWHILE
\IF{$\sum_{k=1}^{j-1} \tilde{p}_k + \sum_{k=j}^i p_k^* < P$}
%\STATE $\bigl\{\tilde{p}_j,\ldots,\tilde{p}_i \bigr\} = \bigl\{p^*_j,\ldots,p_i^* \bigr\}$
\STATE $\forall k \in \{j,\dots,i\},\; \tilde{p}_k \gets p_k^*$
\ELSE
\STATE $\forall k \in \{j,\dots,i\},\; \tilde{p}_k \gets \frac{P - \sum_{k=1}^{j-1}p_k^*}{i-j+1}$
\ENDIF
\STATE $i \gets i+1$
\ENDIF
\ENDWHILE
\RETURN $(\tilde{p}_1,\ldots,\tilde{p}_K)$
\end{algorithmic}
\end{algorithm}
The algorithm is designed the following way. First, allocated power values are initialized to $0$ (line $1$). While allocated power does not exceed $P$, the allocation of power continues (line $3$). If the individual optimal power of mobile user $i$ is strictly less than the individual optimal power of the next mobile user (line $4$),
\begin{itemize}
\item if the sum of the individual optimal power of mobile user $i$ to the power values already allocated does not exceed $P$, this individual optimal power is allocated (line $6$),
\item if the sum of the individual optimal power of mobile user $i$ to the power values already allocated exceeds $P$, only the remaining power is allocated (line $8$). 
\end{itemize}
If the individual optimal power of mobile user $i$ is equal to the individual optimal power of the next mobile user (line $11$), the number of successive equal individual optimal power values is counted (line $13$),
\begin{itemize}
\item if the sum of these optimal power values to the previously allocated power values does not exceed the constraint, these optimal power values are allocated (line $17$),
\item if the sum of these optimal power values to the previously allocated power values exceeds the constraint, the remaining power is fairly shared (line $19$).
\end{itemize}

Note that the algorithm takes into account the case in which several individual optimal power values are equal. Such an event occurs with almost null probability when the channel gains are considered to be continuous and follow an exponential distribution law. 

\begin{figure}
\centering
\includegraphics[width=\linewidth]{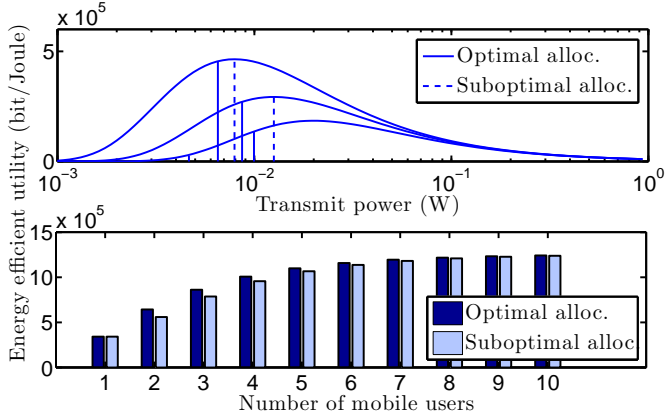}
\caption{Energy-efficient performance comparison of the optimal power allocation and the proposed allocation.}
\label{fig: opti_subopti}
\end{figure}
For the same sum power constraint, Fig. \ref{fig: opti_subopti} illustrates the performance gap between the outcome of the proposed algorithm and the optimal power allocation. The figure on top compares the optimal power allocation and the proposed allocation in a cell with three mobile users. For each user, the associated energy-efficient utility is represented, and the power allocations are given for the optimal and suboptimal cases. While both allocations saturate the sum power constraint, the optimal allocation verifies the slopes equality, in accordance with Sec.~\ref{sec: slopes_prop}, whereas the proposed allocation gives the individual optimal power values to the two mobile users with the best channel gains, and only the remaining power for the third mobile user. The figure below represents the energy efficiency in the cellular cell for the two allocations, for various amount of mobile users in the cell. For each number of mobile users, the channel gains and the sum power constraint are chosen such that the two allocations differ the most. It can be considered as a worst-case scenario for the proposed algorithm.

\section{Selfish Channel State Reporting}\label{sec:game_study}
\subsection{Definition of a game}
This section focuses on the behavior of the mobile users. The main difference with the base station behavior is that mobile users are not concerned with energy efficiency in downlink as they do not provide transmit power themselves. Instead, they are only concerned about their individual SNR. From an operator perspective, the energy consumed at the base station dominates the one needed by the mobile users. Therefore, it makes sense to consider that only the base station is concerned by energy efficiency while the mobile users are only concerned about their SNR.
For each mobile user, this SNR is proportional to the power allocated by the base station. Sec.~\ref{sec: EEschedu} shows that this power depends on the channel gain the considered user reports to the base station, but also on the channel gains reported by of all the other users. Considering that each mobile user prefers to have a high allocated power, it can try to twist the channel gain it reports to the base station in order to get higher allocated power. Hence, we assume that the mobile users have the freedom of sending whatever channel gain feedback they want to the base station, and we use game theory to study what are the outcomes of such a scenario. For each mobile user $k \in \mathcal{K}$, two values of the channel gain are important: 
\begin{itemize}
 \item its actual channel gain $|h_k|^2$, as the actual performance of the mobile user depends on this gain;
\item the value it reports to the base station that is denoted by $g_k \in [0,G]$, with $G$ the maximum gain a mobile user can report. Power is allocated to the mobile user based on this value.
\end{itemize}

Both terms appear in the utility of each mobile user, which is the SNR after power allocation by the base station. $\forall k \in \mathcal{K}$,
\begin{equation}
 u_k (g_k,g_{-k})= \frac{\tilde{p}_k(g_k,g_{-k}) |h_k|^2}{\sigma^2}.
\end{equation}
With a slight abuse of notation $(g_k,g_{-k})$ emphasizes the feedback of mobile user $k$ compared to the CSI reports of all the other mobile users.
We can now properly define the game under study.
\begin{definition}
The channel feedback game is defined by the triplet $\mathcal{G} = (\mathcal{K}, \mathbb{R}, \{u_k\}_{k \in \mathcal{K}})$ in which
\begin{itemize}
 \item $\mathcal{K}$ is the set of players of the game, which represents the mobile users.
\item $\mathcal{A}_k = [0,G]$ is the set of actions for player $k$. In this game, an action $g_k \in\mathcal{A}_k $ is the report to the base station. The set of actions profiles is denoted by $\mathcal{A}=\times_{k = 1}^K \mathcal{A}_k$, and the cardinal product of the actions sets of all players except player $k$ is denoted by $\mathcal{A}_{-k} = \times_{i \neq k} \mathcal{A}_i$.
\item The utility of player $k$ is
\[
  u_k (g_k,g_{-k})= \frac{\tilde{p}_k(g_k,g_{-k}) |h_k|^2}{\sigma^2}.
\]
It is its SNR given the power allocated by the base station knowing all the reported channel gains.
\end{itemize}
\end{definition}

% In this context, the power allocation of the base station naturally depends on the channel gains with every mobile users. But the base station needs the mobile users to report these gains themselves. What would happen if the users were given some degree of freedom in the choice of their report? This is precisely what we want to study via a game theoretical viewpoint and, instead of reporting $h_k$, we assume that mobile users report $\tilde{h}_k$ in order to maximize the transmit power from the base station. 

\subsection{Characterization of the Nash Equilibrium}
Generally, an important concept to study the outcome of a game is the Nash equilibrium.
\begin{definition}
In the game $\mathcal{G}$, a Nash equilibrium is an action profile $(g_1^*,\ldots,g_K^*)$ such that $\forall k \in \mathcal{K}$
\begin{equation}
 \forall g_k \neq g^*_k,\; u_k(g_k, g^*_{-k}) \leq u_k(g_k^*, g^*_{-k}).
\end{equation}
\end{definition}
In other words, it is an action profile from which no player has interest to deviate unilaterally. It is a selfish equilibrium. Interestingly, in the game $\mathcal{G}$, we can prove that there exists at least a pure Nash equilibrium. 
\begin{figure}
\centering
\includegraphics[width=\linewidth]{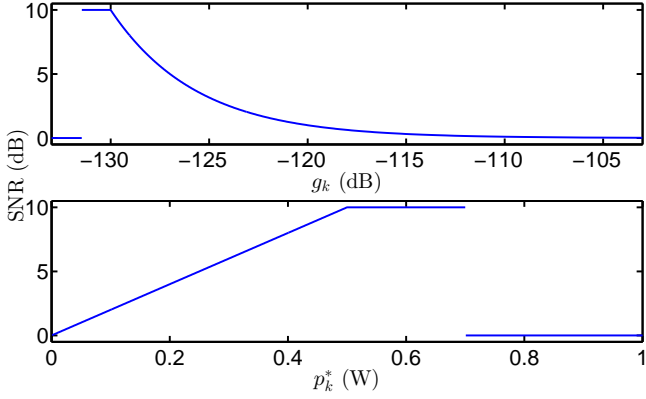}
\caption{SNR of player $k$ with respect to the reported channel gain and its associated individual optimal power.}
\label{fig: uti_game}
\end{figure}

%\begin{definition}{Reciprocal upper semicontinuity~\cite{Simon_1987}}
%\newline The game $\mathcal{G}$ is reciprocally upper semicontinuous if, $\forall i \in \mathcal{K}$, for every sequence $(\mathbf{g}_n)_{n \in \mathbb{N}} \in \mathcal{A}$ such that 
%\begin{equation}
%\lim_{n \to \infty} u_i(\mathbf{g}_n) >  u_i(\lim_{n \to \infty} \mathbf{g}_n),
%\end{equation}
%there exists $j \in \mathcal{K}$ and a sequence $(\mathbf{g'}_n)_{n \in \mathbb{N}}\in \mathcal{A}$ such that
%\begin{equation}
%\lim_{n \to \infty} u_j(\mathbf{g'}_n) <  u_j(\lim_{n \to \infty} \mathbf{g'}_n),
%\end{equation}
%\end{definition}
%
%\begin{proof}
%If the sum of the payoffs is continuous, then the game is reciprocally upper semicontinuous. \red{To be developped.}
%\end{proof}
%
%\begin{definition}{Payoff security.}
%\newline A player $k \in \mathcal{K}$ can secure a payoff of $\alpha \in \mathbb{R}$ at $\mathbf{g} \in \mathcal{A}$ if there exists $\bar{g}_k \in \mathcal{A}_k$, such that $u_k(\bar{g}_k,g_{-k}') \geq \alpha$, $\forall g_{-k}'$ in some open neighborhood of $g_{-k}$.
%A game is payoff secure if $\forall \mathbf{g} \in \mathcal{A}$ and $\forall \epsilon > 0$, each player $k$ can secure a payoff of $u_k(\mathbf{g}) - \epsilon$ at $\mathbf{g}$.
%\end{definition}

\begin{proposition}
 There is a unique Nash equilibrium in the game $\mathcal{G}$. It is $\mathbf{g}^*=(g^*,g^*,\ldots, g^*)$ with $g^*= \frac{K a \sigma^2}{P}$ such that
\begin{equation}
 \arg \max_p \frac{f(\gamma(p,g^*))}{p} = \frac{P}{K}.
\end{equation}
At this equilibrium, power is uniformly allocated among the mobile users and the SNR of player $k \in \mathcal{K}$ is 
\begin{equation}
u_k(\mathbf{g}^*) = \frac{P |h_k|^2}{K \sigma^2}.
\end{equation}
\end{proposition}
It is interesting to note that this equilibrium depends only on the number of mobile users in the cell, and the sum power constraint. The actual channels gains are absolutely not involved in the expression of this equilibrium.

\begin{proof}
First, we show that the action profile $(g^*, g^*, \ldots, g^*)$ is a Nash equilibrium.
Assume that player $k$ deviates from the action $g^*$, it chooses an action $g_k \neq g^*$. The action profile is then $(g_k, g^*_{-k})$.
\begin{itemize}
 \item If $g_k < g^*$ then $p^*_k > \frac{P}{K}$ , and 
\begin{equation}
\sum_{i \in \mathcal{K}} p_i^* = (K-1)\frac{P}{K} + p_k^*>P.
\end{equation}
Hence, according to the allocation algorithm, $\tilde{p}_k$ is set to $P - (K-1)\frac{P}{K} = \frac{P}{K}$, i.e., the SNR of player $k$ remains unchanged.
\item If $g_k > g^*$ then $p_k^*  < \frac{P}{K}$ , then 
\begin{equation}
\sum_{i \in \mathcal{K}} p_i^* = (K-1)\frac{P}{K} + p_k^* < P.
\end{equation}
In this case, all the players are given the individual optimal power. Hence player $k$ gets $\tilde{p}_k = p_k^* < \frac{P}{K}$, and its SNR decreases.
\end{itemize}
We have proven that $\forall k \in \mathcal{K}$, player $k$ has no interest in deviating from the power profile $(g^*, g^*, \ldots, g^*)$, hence it is a Nash equilibrium.
We now prove that this equilibrium is unique.
\begin{itemize}
 \item If the sum of individual optimal power values is below the sum power constraint, i.e., 
\begin{equation}
 \sum_{k \in \mathcal{K}} p_k^* < P,
\end{equation}
then if one player $k$ decreases $g_k$, it gets a higher $\tilde{p}_k$. Hence there can be no equilibrium when the maximum total power constraint is not active.
\item If the sum of individual optimal power values is higher than the sum power constraint, and $\exists k \in \mathcal{K}$ such that $\tilde{p}_k = 0$. Then player $k$ can increase its report $g_k$ in order to have some power allocated. Hence player $k$ has interest in deviating.
\item If the sum of individual optimal power values is higher than the sum power constraint, if no player gets zero power allocated, and $\exists (i,j) \in \mathcal{K}^2$ such that $\tilde{p}_i < \tilde{p}_j$, then player $i$ can slightly decrease $g_i$ in order to get more power from the base station.
\item If the sum of individual optimal power value is higher than the sum power constraint, and $\forall  (i,j) \in \mathcal{K}^2$, $\tilde{p}_i = \tilde{p}_j$, then $\forall i \in \mathcal{K}$, $\tilde{p}_i =  \frac{P}{K}$. If $\exists i \in \mathcal{K}$ such that $p_i^* > \frac{P}{K}$, then any other player $j \neq i$ can report $g_j$ such that $p_j^* \in ]\frac{P}{K},p_i^*[$. In that case, player $j$ receives more power from the base station.
\end{itemize}
We have proven that there cannot be any other equilibrium than the case for which the sum of individual optimal power values saturates the sum power constraint, and $\forall i \in \mathcal{K}$, $\tilde{p}_i = p_i^*=  \frac{P}{K}$.
\end{proof}

Given that at this equilibrium, mobile users do not report their actual channel gains, there are two energy-efficient utilities per cell of interest. First, the energy-efficient utility per cell the base station is convinced to have, given the equilibrium reports of the mobile users. This is not the actual value of the energy-efficient utility per cell. It writes
\begin{equation}
\begin{aligned}
 u_{BS}(\frac{P}{K},g^*) &= \sum_{k \ in \mathcal{K}} \frac{K}{P} \exp(-\frac{a K \sigma^2}{P g^*}), \\
&= \frac{K^2}{P} \exp(-1).
\end{aligned}
\end{equation}
Once again, we can check that this utility depends only on the total number of mobile users in the cell and the power constraint.
The second energy-efficient utility per cell of interest is of course the actual energy efficiency of the cell, which takes into account the true channel gains. This utility writes
\begin{equation}
 u_{BS}(\frac{P}{K}, |h_k|^2) = \frac{K}{P} \sum_{k \in \mathcal{K}} \exp(-\frac{g^*}{|h_k|^2}).
\end{equation}
Hence, the ratio between the actual energy-efficient performance of the cell and the believed performance is
\begin{equation}
\frac{  u_{BS}(\frac{P}{K}, |h_k|^2) }{u_{BS}(\frac{P}{K},g^*)} = \frac{1}{K} \sum_{k \in \mathcal{K}} \exp(1-\frac{g^*}{|h_k|^2}).
\end{equation}

\section{Numerical results}\label{sec: Numeric}

\begin{figure}
\centering
\includegraphics[width=\linewidth]{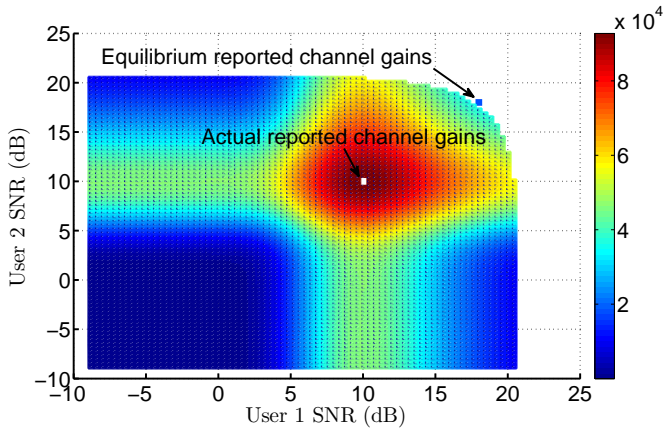}
\caption{Energy-efficient utility of the cell with respect to the SNR of users $1$ and $2$. The color bar represents the cell energy-efficient utility in bit/Joule. Parameters are $a=10$, $|h_1|^2 = |h_2|^2 = -112$ dB, $P=1$ W, and $\sigma^2 = 5\times 10^{-14}$ W.}
\label{fig: 2players}
\end{figure}

In Fig.~\ref{fig: 2players}, a scenario with two mobile users is considered. For both of them, the channel gains are set to $-112$ dB, and we study how the energy-efficient utility per cell varies for all the possible combinations of allocated power values. Hence, the energy-efficient utility per cell is represented with respect to SNR of each of the mobile users. The constraint on the total transmit power is $P = 1W$. On the one hand, we can check that the energy efficiency of the cell is maximized when mobile users report their actual channel gain. But at this point, the sum power constraint is not saturated. Hence, it is not Pareto-optimal for the mobile users. On the other hand, when the mobile users report the equilibrium channel gains, a Pareto optimal point is reached. But in this configuration, the energy efficiency of the cell is not a maximum.

\begin{figure}
\centering
\includegraphics[width=\linewidth]{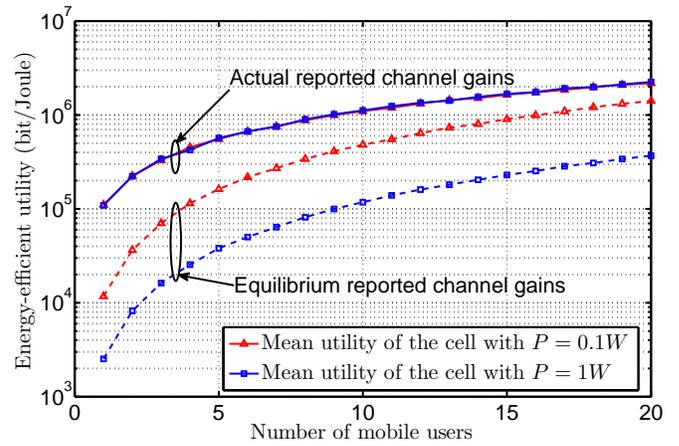}
\caption{Energy-efficient utility of the cell with respect to the number of mobile users.}
\label{fig: Kplayer_BS}
\end{figure}

In Fig.~\ref{fig: Kplayer_BS}, the energy-efficient utility of the cell is represented as a function of the number of mobile users in the cell. The variance of the noise, $\sigma^2$ is set to $5 \times 10^{-14}$ W. The parameter $a$ is set to $6$. Channel gains are assumed to be exponentially distributed. For each number of mobile users, $10^4$ realizations are computed, and the presented results are the mean over these realizations. Similarly to Fig.~\ref{fig: 2players}, we compare the case in which mobile users actually report their channel gains, and the case in which they twist their reports in order to maximize their own utilities. In addition, two power constraints are considered $P=0.1$ W, and $P=1$ W. With no surprise, for both power constraints, the energy-efficient utility of the cell is higher when mobile users report their actual channel gains. This utility increases with the number of mobile users simply because it is a sum over the mobile users. Interestingly, when mobile users report their actual channel gains, there is no significant gap between the two power constraints. There are two explanations for this phenomenon. First, when the sum power constraint is not saturated, the power allocation is exactly the same whatever the power constraint. Second, we recall that mobile users with bad channel gains are not interesting in terms of energy efficiency as their individual optimal power is high and the associated energy efficiency is low. Hence, allocating power or not to these mobile users does not make a significant difference in terms of energy efficiency.
Another interesting phenomenon is when mobile users report equilibrium channel gains, the energy-efficient utility of the cell is worse for $P=1$ W than for $P=0.1$ W. We recall that at the equilibrium, the sum power constraint is saturated. Hence, there is more power allocated to the mobile users in the case with $P=1$ W. But after a certain threshold, increasing power decreases the energy efficiency of the overall system. 

\begin{figure}
\centering
\includegraphics[width=\linewidth]{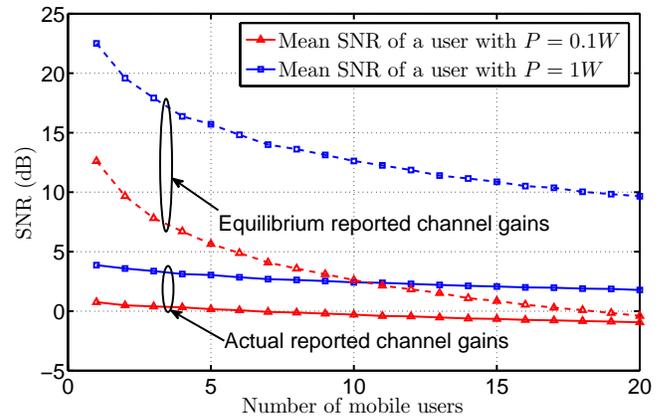}
\caption{Mean SNR of one mobile user with respect to the number of mobile users.}
\label{fig: Kplayer_player}
\end{figure}

In Fig.~\ref{fig: Kplayer_player}, the mean SNR of a mobile user is represented with regard to the number of mobile users for the same scenario as Fig.~\ref{fig: Kplayer_BS}. The SNR of the mobile user is given in dB, and for both power constraints, this SNR is higher when mobile users report equilibrium channel gains. As the SNR of the mobile user at the equilibrium is proportional to the sum power constraint, we can check there is a gain between the two sum power constraints. When mobile users report their actual channel gains, their SNR is lower. But a higher sum power constraint allows the base station to serve more users. Thus the mean SNR of the mobile users with actual channel gains increases with the sum power constraint.

\section{Conclusion}
In this paper we studied the conflict of interest between (i) mobile users that selfishly choose their CSI reports to increase their SNR and (ii) a base station that aims to maximize the cell's energy efficiency by power allocation. We formulated power allocation as a non-convex optimization problem, stated its optimality conditions, and derived a low complexity algorithm. Formulating selfish CSI reporting as a game allowed us to prove the existence of a unique equilibrium from which no user has interest to deviate. Consequently, we obtained a stable system for which we derived energy efficiency with and without true CSI in closed form. 

We illustrate this powerful theoretical framework by numerical results. Naturally, these results show that selfish CSI reports allow mobile users to increase their own SNR while the cell's energy efficiency decreases. Interestingly, this degradation diminishes for smaller sum power constraints and for a larger number of users. Consequently, small cells are more robust to selfish CSI reports than cells with large power constraints and few users.

As future work, we aim to extend this framework by an optimal power allocation algorithm, energy-efficient utilities not only for the base station but for the mobile users, as well as by a practical, discrete set of CSI values.

%Contribution of this paper:
%\begin{itemize}
%\item Energy-efficient power allocation
%\begin{itemize}
%\item Conditions for optimality
%\item Simple suboptimal algorithm with decent performance
%\end{itemize}
%\item Game study when transmitters do not report their actual channel gains
%\begin{itemize}
%\item Characterization of the Nash equilibrium
%\item Performance at this equilibrium
%\end{itemize}
%\end{itemize}
%
%Perspectives:
%\begin{itemize}
%\item Optimal power allocation algorithm
%\item Game with energy-efficient utilities for the mobile users
%\item Discrete set of channel gains reports
%\end{itemize}

%\begin{acronym}
    \acrodef{PPT}{point-to-point}
    \acrodef{M2M}{Machine-to-Machine}
    \acrodef{BER}{Bit Error Rate}
    \acrodef{16-QAM}{16 Quadrature Amplitude Modulation}
    \acrodef{64-QAM}{64 Quadrature Amplitude Modulation}
    \acrodef{256-QAM}{256 Quadrature Amplitude Modulation}
    \acrodef{ACF}{Autocorrelation Function}
    \acrodef{ARQ}{Automatic Repeat Request}
    \acrodef{aDA}{advanced Dynamic Algorithm}
    \acrodef{ADC}{Analog-to-Digital Converter}
    \acrodef{APP}{Application layer}
    \acrodef{ASIC}{Application Specific Integrated Circuits}
    \acrodef{AWGN}{Additive White Gaussian Noise}
    \acrodef{BPSK}{Binary Phase Shift Keying}
    \acrodef{CBR}{Constant Bit Rate}
    \acrodef{CC}{Chase Combining}
    \acrodef{CDF}{Cumulative Distribution Function}
    \acrodef{CDMA}{Code-Division-Multiple-Access}
    \acrodef{CIC}{Cascaded Integrator-Comb}
    \acrodef{CIF}{Common Intermediate Format}
    \acrodef{CNR}{Channel Gain-to-Noise Ratio}
    \acrodef{CRC}{Cyclic Redundancy Check}
    \acrodef{CSI}{Channel State Information}
    \acrodef{CS}{Carrier Sensing}
    \acrodef{DAC}{Digital-to-Analog Converter}
    \acrodef{DCT}{Discrete Cosine Transform}
    \acrodef{DIV}{Distortion In Interval}
%!! don't use DLC acronym in this paper (it's defined in the intro, simply use "DLC" instead)
%    \acrodef{DLC}{Data Link Control layer}
    \acrodef{DSP}{Digital Signal Processor}
    \acrodef{EGC}{Equal Gain Combining}
    \acrodef{FDMA}{Frequency Division Multiple Access}
    \acrodef{FEC}{Forward Error Correction}
    \acrodef{FER}{Frame Error Rate}
    \acrodef{FS}{Frame Selection}
    \acrodef{FIFO}{First-In-First-Out}
    \acrodef{FPGA}{Field Programmable Gate Array}
    \acrodef{FSC}{Frame Check Sequences}
    \acrodef{GMSK}{Gaussian Minimum Shift Keying}
    \acrodef{GoP}{Group of Pictures}
    \acrodef{GSR}{GNU Software Radio}
    \acrodef{HTTP}{Hypertext Transfer Protocol}
    \acrodef{HTML}{Hypertext Mark-up Language}
    \acrodef{ICI}{Inter-carrier Interference}
    \acrodef{IEEE}{Institute of Electrical and Electronics Engineers, Inc.}
    \acrodef{IP}{Internet Protocol}
    \acrodef{IPC}{Inter-Process Communication}
    \acrodef{ISI}{Inter-symbol Interference}
    \acrodef{ISM}{industrial, scientific and medical}
    \acrodef{LLC}{Logical Link Control layer}
    \acrodef{LOS}{Line Of Sight}
    \acrodef{MAC}{Medium Access Control}
    \acrodef{MCM}{Multi Carrier Modulation}
    \acrodef{MIMO}{Multiple-Input Multiple-Output}
    \acrodef{MISO}{Multiple-Input Single-Output}
    \acrodef{MPEG}{Moving Pictures Expert Group}
    \acrodef{MOS}{Mean Opinion Score}
    \acrodef{MRC}{Maximum Ratio Combining}
    \acrodef{SC}{Selection Combining}
    \acrodef{MSB}{Most Significant Bit}
    \acrodef{MSS}{Maximum Segment Size}
    \acrodef{MTU}{Maximum Transmission Unit}
    \acrodef{NAV}{Network Allocation Vector}
    \acrodef{NCR}{Non-Cooperative Relaying}
    \acrodef{NLOS}{Non-Line Of Sight}
    \acrodef{NPS}{Network Path Selection}
    \acrodef{OFDM}{Orthogonal Frequency Division Multiplexing}
    \acrodef{OS}{Operating System}
    \acrodef{OR}{Opportunistic Relaying\slash{}Routing}
    \acrodef{PDF}{Probability Density Function}
    \acrodef{PDU}{Protocol Data Unit}
    \acrodef{PER}{Packet Error Rate}
    \acrodef{PHY}{Physical layer}
    \acrodef{PSC}{Packet Selection Combining}
    \acrodef{PSNR}{Peak Signal-to-Noise Ratio}
    \acrodef{QCIF}{Quarter CIF}
    \acrodef{QoS}{Quality of Service}
    \acrodef{QPSK}{Quadrature Phase Shift Keying}
    \acrodef{RCPC}{Rate-Compatible Punctured Convolutional}
    \acrodef{RF}{Radio Frequency}
    \acrodef{RMS}{root mean square}
    \acrodef{RTT}{Round Trip Time}
    \acrodef{SDF}{Selection Decode-and-Forward}
    \acrodef{SDR}{Software Defined Radio}
    \acrodef{SR}{Software Radio}
    \acrodef{SEP}{Symbol Error Probability}
    \acrodef{SCM}{Single Carrier Modulation}
    \acrodef{SDC}{Selection Diversity Combining}
    \acrodef{SIFS}{Short Inter-Frame Space}
    \acrodef{SNR}{Signal-to-Noise Ratio}
    \acrodef{TCP}{Transmission Control Protocol}
    \acrodef{TDMA}{Time Division Multiple Access}
    \acrodef{UDP}{User Datagram Protocol}
    \acrodef{USRP}{Universal Software Radio Peripheral}
    \acrodef{VBR}{Variable Bit Rate}
    \acrodef{VQM}{Video Queue Management}
    \acrodef{WLAN}{Wireless Local Area Network}
    \acrodef{WMAN}{Wireless Metropolitan Area Network}
    \acrodef{WSN}{Wireless Sensor Network}
    \acrodef{WT}{Wireless Terminal}
    \acrodef{WWW}{World-Wide-Web}
%\end{acronym}

\bibliographystyle{abbrv}
\bibliography{JRNLabrv,ConfAbrv,IEEEabrv,belllabs}  % sigproc.bib is the name of the Bibliography in this case
% ACM needs 'a single self-contained file'!
%\balancecolumns

%\balancecolumns % GM June 2007
% That's all folks!
\end{document}